# Fingering instability in Marangoni spreading on a deep layer of polymer solution

Xue Ma[1], Menglin Zhong[1], Yifeng He[1], Zhanwei Liu[1], Zhenzhen Li[*,1]

[1]School of Aerospace Engineering, Beijing Institute of Technology, 100081 Beijing, P. R. China.

Correspondence and requests for materials should be addressed to ZZ.L. (email: zhenzhenli@bit.edu.cn)

**Abstract**

Spreading on the free surface of a complex fluid is ubiquitous in nature and industry, owing to the wide existence of complex fluids. Here we report on a fingering instability that develops during Marangoni spreading on a deep layer of polymer solution. In particular, the wavelength depends on molecular weight and concentration of the polymer solution. We use the Transmission Lattice Method to characterize the finger height at the micron scale. We model the evolution of spreading radius, involving viscoelastic and shear thinning effects, to suggest a more generalized law than the spreading of Newtonian fluids. We give physical explanation on the origin of the fingering instability as due to normal stresses at high shear rate generating high contact angle and deformation at the leading edge, and so selects the wavelength of the fingering instability. Understanding the spreading mechanism has particular implication in airway drug delivery, and surface coating with patterns.

Marangoni spreading has wide application in nature and industry, such as oil spreading on the sea[1], motion of living organisms[2], ultra clean drying process[3], encapsulation process[4], coating of functional films[5], and visualizing the shortest path line in maze problems[6]. Particularly, Marangoni flow is involved in airway and pulmonary disease treatment, such as Surface Replacement Therapy[7], and delivery of antibiotics for airway infection[8], combined with inhalation of atomized drug components[9]. Since the presence of mucin renders the airway surface viscoelastic nature[10], normal stress plays important role in affecting surface deformation and surfactant concentration distribution during Marangoni spreading[11]. However the effect of viscoelasticity on the leading edge instability of Marangoni spreading has not been mentioned to our best knowledge, although viscoelasticity has been shown to play an important role in other types of instability, such as that of a thin fluid layer flowing along an inclined surface[12,13], viscous fingering[14], and contact line instability[15].

The fingering instability near the leading edge of Marangoni spreading is extensively studied for Newtonian fluids on thin film flow[16]. Finger patterns and spreading dynamics are shown to depend on film thickness[17,18], surfactant concentration[19], capillarity and surface diffusion[20], solubility of surfactants and micelles formation[21,22], and evaporative components in the spreading drop[23,24,25]. Theoretical and numerical work arrived to capture the transient behavior of finger formation[26,27]. Despite the fingering instability occurred for Newtonian fluids spreading on a thin film, it has not been to our best knowledge reported on the deep layer Marangoni spreading[28,29].

Here we show that the fingering instability can occur during Marangoni spreading on the free surface of a deep layer of polymer solution, possessing both shear thinning and viscoelastic properties. The leading edge where develops a fingering instability can be well identified due to a high apparent contact angle. Spreading radius, free surface height perturbation, and instability wavelength selection are tightly affected by viscoelastic and shear thinning characteristics.

## Results

**Observations.** For showing the importance of shear thinning and viscoelastic effects in the fingering instability, side views of three experiments are recorded for comparison. First experiment, a high molecular weight polymer solution (PEO Mw=5 MDa) without surfactants is dropped onto a deep layer of the same polymer solution (Supplementary Fig. 1a and Movie 1), which is shear thinning and viscoelastic. A typical merging procedure is observed, with a delayed merging[30] for the first 18 ms, before complete merging; no instability is observed. Capillary waves propagate up along the drop-air interface, but did not lead to complete pinching of the drop as in a Newtonian fluid case[31]. This is due to the higher characteristic pinching time of polymer filament compared with the characteristic time of wave propagation. The time for the capillary wave to reach the top is[31] $t_c = \pi r / (\gamma k/\rho)^{1/2}$, with $r$ the drop radius = 1.5 mm, $\gamma$ the drop surface tension = 31.5 mN/m, $k$ the wavenumber, and $\rho$ the fluid density = $10^3$ kg/m³. For capillary wave which has the wavelength $\lambda_c$ smaller than the capillary length[32] $\kappa^{-1} = \sqrt{\frac{\gamma}{\rho g}} = 1.8$ mm, the wavenumber $k = \frac{2\pi}{\lambda_c} > 3488$. So that $t_c < 14.2$ ms. Whereas the characteristic time for polymer filament pinching is typically several tens or hundreds of ms for our tested samples (Supplementary Table 1). Second experiment, Marangoni spreading of a low molecular weight polymer solution (Supplementary Fig. 1b and Movie 1), PEO solution (Mw=0.3 MDa) with 10 cmc surfactant in the drop is dropped onto a deep layer of the same PEO solution without surfactant. This PEO solution has constant viscosity with varying shear rate, therefore can be considered as a Newtonian fluid. A typical Marangoni spreading is observed, with a Reynold ridge propagating radially outwards[28]; no instability is observed. Third experiment, Marangoni spreading of a high molecular weight polymer solution (Fig. 1a and Movie 1), PEO (Mw=5 MDa) with 10 cmc surfactant is dropped onto a deep layer of the same polymer solution without surfactant. An instability occurs at the leading edge since the very beginning of spreading, and evolves to form large and flat fingers. The fingers are slightly higher than the bath at its vicinity. This comparison shows that a shear thinning and viscoelastic fluid exhibits a fingering instability during Marangoni spreading on a deep layer, while a Newtonian fluid does not in the similar conditions. This indicates that a high enough surface flow velocity and high enough molecular weight (Mw) are the two necessary conditions for having fingers. Although the drop and the bath are miscible liquids, fast spreading of high Mw solution creates a distinguishable leading edge, on which the fingers develop.

The fingers can be observed in a vertical view by a transmission and reflection optical method (Fig. 1b and Movie 2). After the drop deposition, the surface waves propagate the fastest, followed by the leading edge with the fingering instability. This difference in propagation velocity is reminiscent of a

Newtonian fluid spreading on a liquid bath[33]. The vertical view allows measurements on the spreading dynamics, including the spreading radius, the leading edge velocity and the wavelength of fingers that will be shown and discussed later.

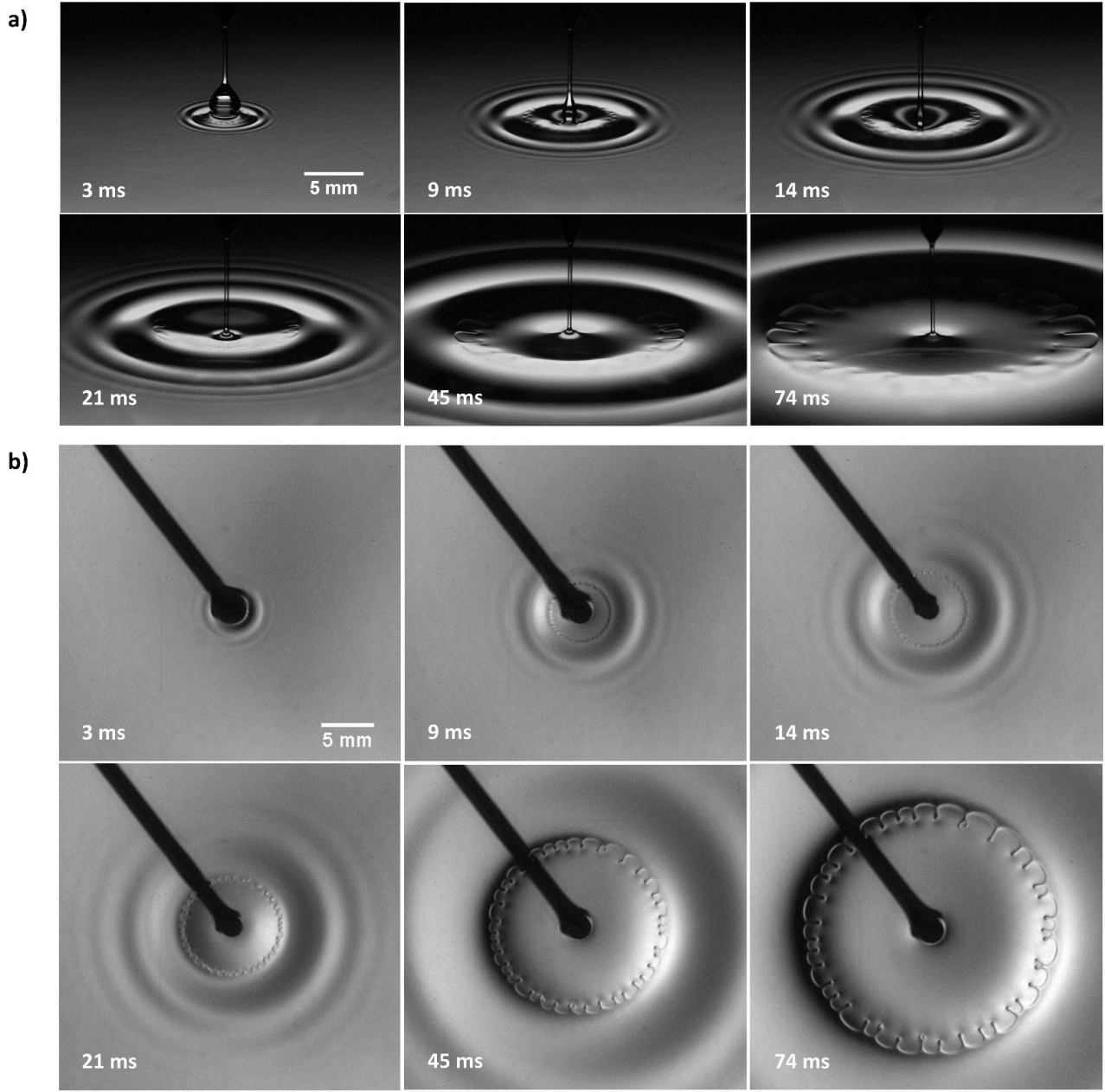

**Figure 1: Marangoni spreading on a free liquid surface on the deep layer of a shear thinning and viscoelastic fluid. a,** Side view of a drop of PEO solution (Mw=5 MDa, C=4 g/L) with 10 cmc of Triton X-100 non-ionic surfactant spreading on a deep layer (8 mm) of the same PEO solution without surfactant. The fingering instability is observed at the drop leading edge. **b,** Top-down view by a transmission method. Surface waves propagate outwards, followed by a leading edge of spreading on which the fingering instability develops. The time 0 corresponds to the moment when the drop touches the free surface.

**Spreading radius.** Before analyzing the finger formation, we firstly studied the evolution of the radius of the leading edge. Result of PEO solutions (Mw=5 MDa) at different concentrations and with 10 cmc surfactant is shown in Fig. 2, results of PEO solutions with other two molecular weights are shown in Supplementary Fig. 3 & 4. The spreading radius of PEO solution follows power law (Fig. 2a), the velocity of the leading edge is deduced by differentiating the spreading radius with respect to time (Fig. 2b).

In our case of a shear thinning fluid, viscosity is time dependent, since in this transient procedure of spreading, the shear rate is time dependent. We measure the dependence of viscosity $\eta$ on shear rate $\dot{\gamma}$ with a rheometer, and fitted it by the Cross model[34]: $\eta(\dot{\gamma}) = \eta_\infty + \frac{\eta_0 - \eta_\infty}{1+(k\dot{\gamma}(t))^p}$, with $\eta_0$ and $\eta_\infty$ the two plateau values at low and high shear rate, respectively, and coefficient $k$ and $p$ as fitting parameters (Fig. 2c). In order to have the time dependency of shear rate, we suppose that $V$ is the velocity of spreading and $h$ the depth of viscous diffusion over time $\sim \sqrt{\frac{\eta}{\rho}t}$, with $\rho$ the fluid density. We inject the Cross model of viscosity to the expression of $h$, which is injected into the definition of the shear rate $\dot{\gamma}(t) = \frac{V(t)}{h(t)}$, this gives the following equation of $\dot{\gamma}(t)$:

$$\dot{\gamma}(t)^2 \cdot \left(\eta_\infty + \frac{\eta_0 - \eta_\infty}{1+(k\dot{\gamma}(t))^p}\right) = \frac{\rho \cdot V(t)^2}{t} \quad (1)$$

We substitute the experimental $V(t)$ data (Fig. 2b) into equation (1), thus $\dot{\gamma}(t)$ is obtained by numerically solving the equation, and shown in Fig. 2d. A low concentration solution possesses larger shear rate than high concentration solution, because of higher spreading velocity and thinner thickness of the viscous diffusion layer. We thus deduced the temporal evolution of viscosity according to the dependence of viscosity on shear rate, viscosity increases with time more obviously for high concentration solution, for which the shear thinning effect is more pronounced (Fig. 2e). We also deduced the temporal dependence of the viscous diffusion layer thickness $h(t)$ from the values of $\eta(t)$, as shown in Fig. 2f, $h(t)$ increases during time, and a higher concentration solution has a thicker viscous diffusion layer than lower concentration solution.

The dynamics of the spreading is governed by the balance between surface tension gradient with shear stress: $\frac{S}{R} \sim \eta \frac{V}{h}$, with $S$ the spreading parameter, $R$ the spreading radius. Thus we have the temporal evolution of the radius of the leading edge[1] $R(t) \sim \frac{S^{\frac{1}{2}} \cdot t^{\frac{3}{4}}}{(\eta\rho)^{\frac{1}{4}}}$. By plotting $R(t) \cdot \eta(t)^{1/4}$ as function of $t$, with the values of $\eta(t)$ that we obtained in Fig. 2e, all the curves of $R(t) \cdot \eta(t)^{1/4}$ collapse to 3/4 power law when $t$ is higher than the order of 20 ms (Fig. 2g). This indicates that spreading is governed by balance between Marangoni stress and viscous stress when $t > 20$ ms. With the viscosity varying constantly during time in the case of shear thinning fluids, the quantity $R(t) \cdot \eta(t)^{1/4}$ follows the universal 3/4 power dependence on time, a more generalized form of spreading than Newtonian fluids on a deep layer[1,4,28]. However, there is deviation from the 3/4 power when $t < 20$ ms, when the experimental quantity $R(t) \cdot \eta(t)^{1/4}$ shows a more flattened variation with time (Fig. 2g). This indicates that when $t < 20$ ms, the balance of Marangoni stress with viscous stress cannot explain the flattened zone. The time range of $20\ \mathrm{ms}$ is reminiscent of the viscoelastic relaxation time of polymer chain[35], that $T_z = \frac{\eta R_g^3}{k_B T}$, with $R_g = m^{3/5} a$ the gyration radius of a polymer coil in the diluted solution, $m$ the monomer number of a polymer chain, $a$ the size of a monomer, $k_B$ the Boltzmann constant and $T$ the temperature. In aqueous solution, the characteristic viscoelastic time is $13.3\ \mathrm{ms}$ for a PEO chain with 5 MDa.

In this drop spreading on an 8 mm deep layer, the thickness of the viscous diffusion layer is the order of $h \approx 100\ \mu m$ at the beginning of spreading, as shown in Fig. 2f. Reynolds number is defined as $\mathrm{Re} = \frac{\rho V h}{\eta}$. For comparing the relative importance of inertial effect with elastic effect, we estimate the Elastic number[36], $\mathrm{El} = \frac{Wi}{Re} = \frac{\frac{\tau V}{R}}{\frac{\rho V h}{\eta}} = \frac{\tau \eta}{R \rho h}$, with Weissenberg number $Wi = \frac{\tau V}{R}$, the relative importance between the viscoelastic relaxation time $\tau$ and the characteristic time of the spreading procedure $\frac{R}{V}$.

At the beginning of spreading, with $V \approx 0.5\ \mathrm{m/s}$ the velocity (Fig. 2b), $\eta \approx 10\ \mathrm{mPa.s}$ the shear thinned viscosity (Fig. 2e), $\tau \sim 10$ ms the typical viscoelastic relaxation time of our samples (Supplementary Table 1) and $R \approx 1$ mm (Fig. 2a), $\mathrm{Re}$ is estimated to be the order of 5. This indicates that the inertial effect plays an important role at early time spreading, similar to the spreading of Newtonian drop on a deep layer[29]. The elastic number $\mathrm{El} \sim 1$, indicating that the elastic and inertial effects are comparable at early time of spreading. Considering this fact, in the following, we focus our discussion on the interplay of elastic and viscous effects, and propose an explanation based on the Maxwell model, for the temporal evolution of the spreading radius, which is a more generalized law than that of Newtonian fluids.

The viscoelastic character of the solution can be described by a spring and a dashpot connected in series[37]. The elastic shear stress and viscous stress both equal to surface tension gradient along the radial direction: $\sigma_e = \sigma_\eta = \frac{S}{R}$, with $R = r_e + r_\eta$ the total displacement of the leading edge, $r_e$ and $r_\eta$ are the displacements contributed by elastic and viscous shear effects, respectively. The total shear rate is $\dot{\gamma} = \dot{\gamma}_e + \dot{\gamma}_\eta$. The elastic shear stress $\sigma_e = G\gamma_e = G\frac{r_e}{\sqrt{\eta t/\rho}}$ with $G$ the shear modulus equals to $\eta/\tau$, with $\tau$ the viscoelastic relaxation time[38]. The viscous stress $\sigma_\eta = \eta \dot{\gamma}_\eta = \eta \frac{r_\eta/t}{\sqrt{\eta t/\rho}}$. By equating $\sigma_e$ and $\sigma_\eta$, we obtain $\frac{r_e}{r_\eta} = \frac{\eta}{Gt}$. By equating $\sigma_\eta$ with $\frac{S}{r_e+r_\eta}$, we finally obtain:

$$r_\eta(t) = \frac{S^{1/2} \cdot t^{3/4}}{\rho^{1/4} \eta(t)^{1/4} (\frac{\tau}{t}+1)^{1/2}} \quad \text{and} \quad r_e(t) = \frac{\tau S^{1/2} t^{-1/4}}{\rho^{1/4} \eta(t)^{1/4} \cdot (\frac{\tau}{t}+1)^{1/2}} \quad (2)$$

In case of $\tau \to 0$, the above equation $r_e(t)$ tends to 0, and $r_\eta(t) \cdot \eta(t)^{1/4}$ tends to 3/4 power dependence on time. We obtained the viscoelastic time $\tau$ by fitting the first normal stress dependence on shear rate $N_1(\dot{\gamma})$ measured by a rheometer[14], based on the relation that $N_1 = \Psi_1 \dot{\gamma}^2$, with $\Psi_1 = 2 n k_B T \tau^2$ the coefficient of first normal stress, $n$ the number density of polymer molecules, and $k_B T$ the thermal energy (Supplementary Fig. 6). The values of $\tau$ are listed in Supplementary

Table 1 for different molecular weights and concentrations. We apply the values of $\tau$ to the equation (2), and obtained the respective contribution of elastic and viscous effects on the spreading radius (Fig. 2h). We observe that $r_e(t)$ dominates over $r_\eta(t)$ when $t < \tau$; and the elastic and viscous effects interchanges importance at $t = \tau$, and the viscous effect dominates when t further increases. $r_e(t) + r_\eta(t)$ agrees with the experimentally measured spreading radius. More curves for 5 MDa at other concentrations, and for solutions with other two molecular weights are shown in Supplementary Fig. 9 (1-3). The Maxwell model explains the flattening of $R(t) \cdot \eta(t)^{1/4}$ compared with 3/4 power law at the early stage of spreading, indicating an important contribution of elastic effect.

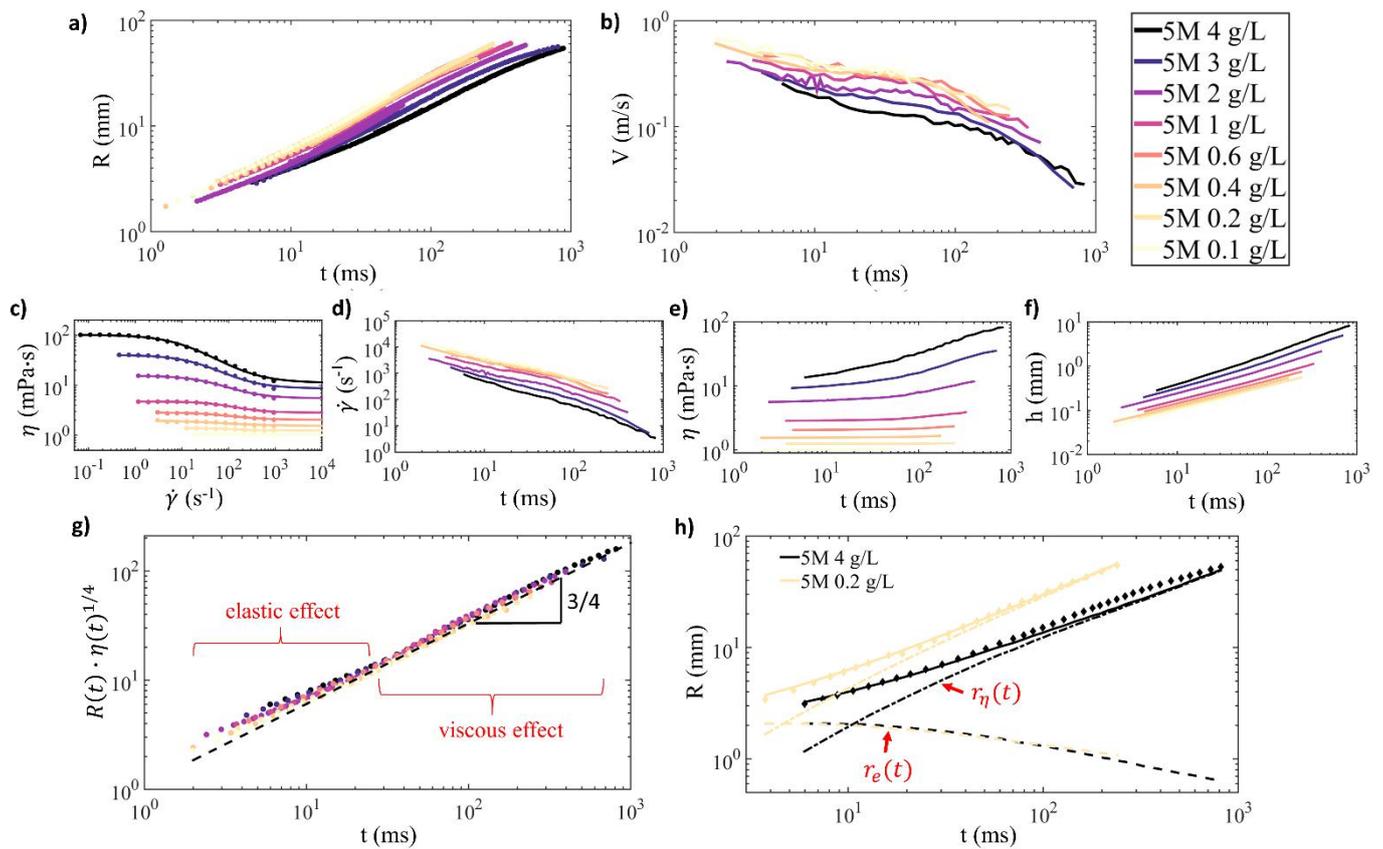

**Figure 2: Spreading radius of PEO drop with surfactant on a deep layer of PEO solution without surfactant. a,** Experimentally measured spreading radius as function of time, for PEO Mw=5 MDa solution at different concentration. Each solution is measured using method of transmission and reflection, both coherent with each other. **b,** Velocity of the leading edge vs. time, deduced from differentiating the radius with respect to time. **c,** Viscosity vs. shear rate measured by a rheometer, and fitted by Cross model. **d,** Shear rate vs. time calculated from equation (1). **e,** Viscosity vs. time for the shear thinning fluids. **f,** Thickness of the viscous diffusion layer vs. time. **g,** Temporal evolution of $R(t) \cdot \eta(t)^{1/4}$ for different concentrations. **h,** The radius measured by experiment (diamond) agrees with the calculated radius $r_e + r_\eta$ from equation (2) (solid lines). Two samples of PEO 5MDa, at concentration 4 g/L (black) and 0.2 g/L (yellow) are shown, dashed lines for elastic displacement $r_e(t)$, and dash-dotted lines for viscous displacement $r_\eta(t)$.

**Height perturbation and visible leading edge.** It can be perceived from Fig. 1a, that fingers are slightly higher than the bulk fluid, the fingering instability is associated with the free surface height elevation. Similar observation occurs in the fingering instability on thin liquid films[20], which was measured by interferometry[23]. We use Transmission Lattice Method to detect the free surface waves and the height perturbation of fingers (Fig. 3). The cross-sectional profile of the free surface is plotted at different time steps (Fig. 3c), showing that the surface wave propagates radially outwards. Propagation of the leading edge and the fingers (marked by dots on the profiles in Fig. 3c) are initially faster than that of surface wave, then falls behind to the crest and finally to the back of the surface wave. We measured the finger height up to 9.5 ms for the sample shown in Fig. 3c, and did not measure for longer time, due to the technical limitation discussed in the Method section. The height perturbation of the fingers above their vicinity are measured along the circumferential direction of the spreading drop (Fig. 3d), thus height of all the fingers $\delta h$ at a fixed time can be determined and averaged. The average height perturbation of the fingers $\overline{\delta h}$ is the order of several microns. At early time of spreading within our measuring time interval, the height of fingers tend to increase (Fig. 3e). However, since the fingers disappear gradually for longer times, we anticipate that the height perturbation of the fingers finally decays with time.

We compare the finger height of three samples with three different molecular weights (2 MDa, 5 MDa, 8 MDa). The concentrations are chosen for the three samples having comparable viscosity (Supplementary Fig. 7), but very different nonlinear elasticity as shown by the first normal stress difference $N_1$ in Fig. 3f. Thus we can illustrate the consequence of nonlinear elasticity on the finger height. $N_1$ is previously measured by a rheometer as function of shear rate, and the temporal evolution of $N_1$ (Fig. 3f) is obtained considering the temporal variation of shear rate (Fig. 2d). Shear rate in Marangoni flow is the highest at the beginning of spreading, and since $N_1 = \Psi_1 \dot{\gamma}^2$ [38], $N_1$ is very important at the beginning of spreading, and then decays rapidly (Fig. 3f). Considering Fig. 3e and 3f, we find that both $N_1$ and the finger height increases with increasing molecular weight, so that higher $N_1$ corresponds to higher average finger height $\overline{\delta h}$, indicating that the free surface height elevation at the fingers is attributable to nonlinear elasticity.

The effect of the nonlinear elasticity at the leading edge can be described by a sketch in Fig. 4. Supposing there is a curvature in $(\vec{e_r}, \vec{e_z})$ plane at the leading edge, this curvature corresponds to a radius $a$. The balance of the stresses in $\vec{e_r}$ leads to equation (3a), with $\frac{\gamma}{a}$ the Laplace pressure created by the curvature, $\gamma$ the surface tension at the leading edge approximated as the surface tension between the drop and the air: 31.5 mPa·s. $\underline{\sigma}^E$ denotes the stress tensor, and $\sigma_{rr}^E$ the elastic normal stress in $\vec{e_r}$ and exerting on a surface with normal direction $\vec{e_r}$. Besides, the profile of the free liquid surface at the upstream of the leading edge is gentle, with negligible curvature. So that the balance of normal stress in $\vec{e_z}$ gives equation (3b). Since $\sigma_{rr}^E - \sigma_{\theta\theta}^E = N_1$ and $\sigma_{\theta\theta}^E - \sigma_{zz}^E = N_2$ with $N_2$ usually negligible compared with $N_1$[38], this leads to equation (3c).

$$-\mathrm{P} + \sigma_{rr}^E = -\mathrm{Pa} + \frac{\gamma}{a} \quad (3a)$$

$$-\mathrm{P} + \sigma_{zz}^E = -\mathrm{Pa} \quad (3b)$$

$$\sigma_{rr}^E - \sigma_{zz}^E \cong N_1 \quad (3c)$$

By combining equation (3a), (3b) and (3c), we have $\frac{\gamma}{a} = N_1$. Since $N_1 > 0$, we have $a > 0$, the meniscus at the leading edge is curved towards $-\vec{e_r}$ with radii of curvature $a$, as shown in Fig. 4a. This leads to a folding of the free surface at the leading edge. Therefore, a relatively high apparent contact angle β at the folding place is formed (Fig. 4a), renders the leading edge well distinguishable. To the contrary, spreading of a Newtonian fluid which does not exhibit nonlinear elastic properties, has hardly distinguishable leading edge. The high contact angle has been shown to be one precondition for forming the contact line instability[15]. In the work of Deblais *et al.*, the high contact angle is imposed by a hydrophobic solid surface on which the polymer film is sheared, leading to a contact line instability. Whereas in our Marangoni spreading on the free surface of a deep layer, the apparent contact angle is created by the first normal stress $N_1$, which is considerable at high shear rate.

To explain the order of magnitude of the curvature at the leading edge, we take the example of Mw=8 MDa and C=2.75 g/L. The $N_1$ at the earliest time of our measuring capacity is that $N_1(t = 2\ ms) \approx 3000\ Pa$, thus the radii of the curvature $a = \frac{\gamma}{N_1} \approx 10\ \mu m$. This is at the same order of magnitude with the height perturbation $\overline{\delta h}$ that we measured at the leading edge (Fig. 3e). With increasing time, $N_1$ decreases dramatically (Fig. 3f), so that the radii of curvature a increases simultaneously. However experimentally the height perturbation $\overline{\delta h}$ does not show a dramatic increase, to the contrary, it increases gently (Fig. 3e) and decays for longer time. This is because of the fact that the apparent contact angle β is initially very high, and gradually decreases as the leading edge becomes invisible. During the same period, the radii of curvature a increases, and go beyond the height perturbation $\overline{\delta h}$. So that although at very beginning the radii of curvature is comparable with the height perturbation of the fingers, they do not have synchronized variation for longer time. The increase of a is synchronized with the decrease of β. Whereas the fact that $\overline{\delta h}$ keeps its magnitude (or vary only gently), can be explained as in a sufficiently accelerated process[39], the strain during an interval of time is determined by the stress and elastic modulus at very early time. Therefore a state of stress during a short time interval in the history leads to a state of strain later for an extended time interval, thus a relative constant $\overline{\delta h}$ during our measurement period. High Mw solution has higher $\overline{\delta h}$ than low Mw solution, this is consistent with wavelength comparison between different Mw, which will be shown in the following.

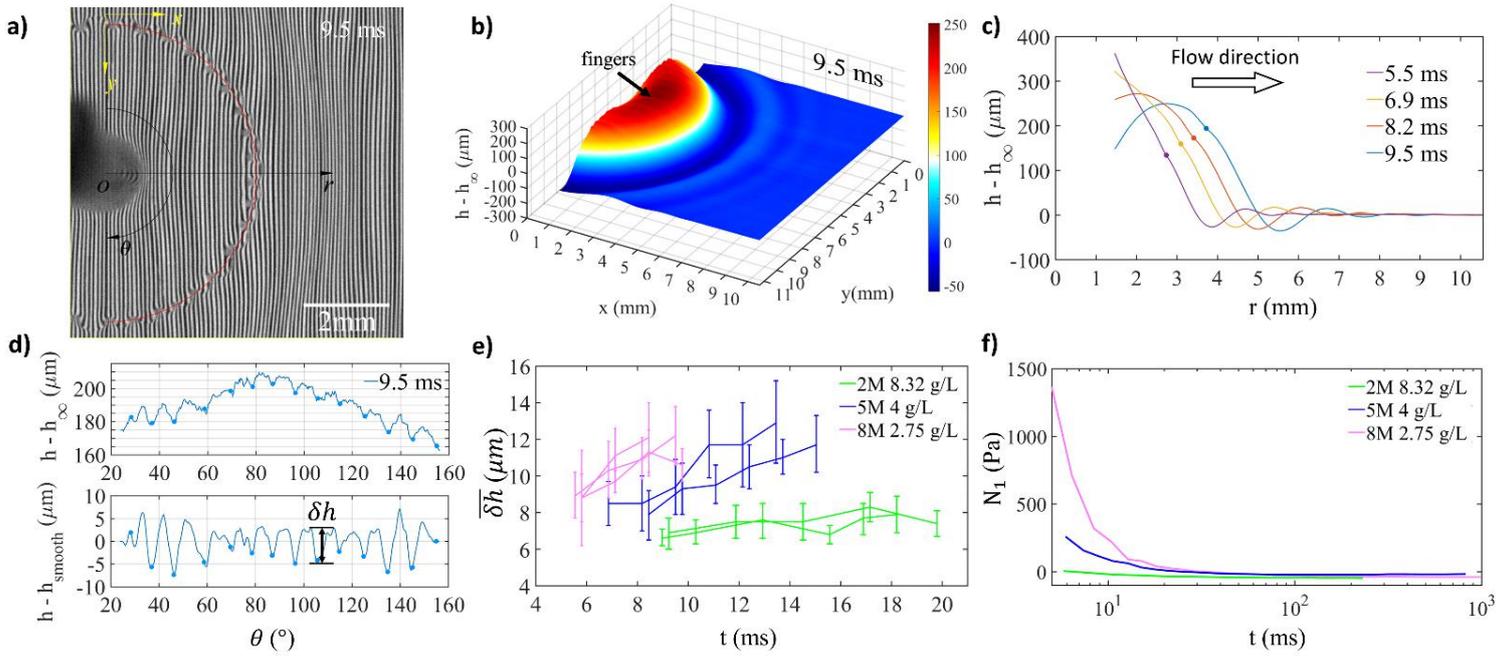

**Figure 3: Free surface deformation induced by Marangoni spreading of PEO solution, measured by Transmission Lattice Method. a,** Deformation of grid patterns caused by fluctuation of liquid surface topography. Red line indicates circumferential coordinate passing through the finger tips. **b,** Free surface topography calculated by Sampling Moiré method, illustrating surface waves, and the darkest red zones near the crest is where occurs the fingers at 9.5 ms. **c,** Cross-sectional profile of the free surface along $\vec{r}$ direction in **a**, at different time steps, for PEO solution at Mw=8 MDa, C=2.75 g/L. $h$ is the free surface height, $h_\infty$ the height of the unperturbed surface at infinity. Points on each curve represent radial position of the finger tips. **d,** Free surface height versus curvilinear position along red line in **a**, at time 9.5 ms. Upper figure for $h - h_\infty$, and this curve is smoothed to get $h_{smooth}$; lower figure for local finger height above the wave: $h - h_{smooth}$, $\delta h$ denotes the height of a finger. **e,** Temporal evolution of the average finger height $\overline{\delta h}$ for three solutions with different molecular weights (2 MDa, 5 MDa, 8 MDa), at concentrations with similar viscosity. **f,** Temporal evolution of the first normal stress difference $N_1$ measured by a rheometer and calculated according to $\dot{\gamma}(t)$ obtained from equation (1).

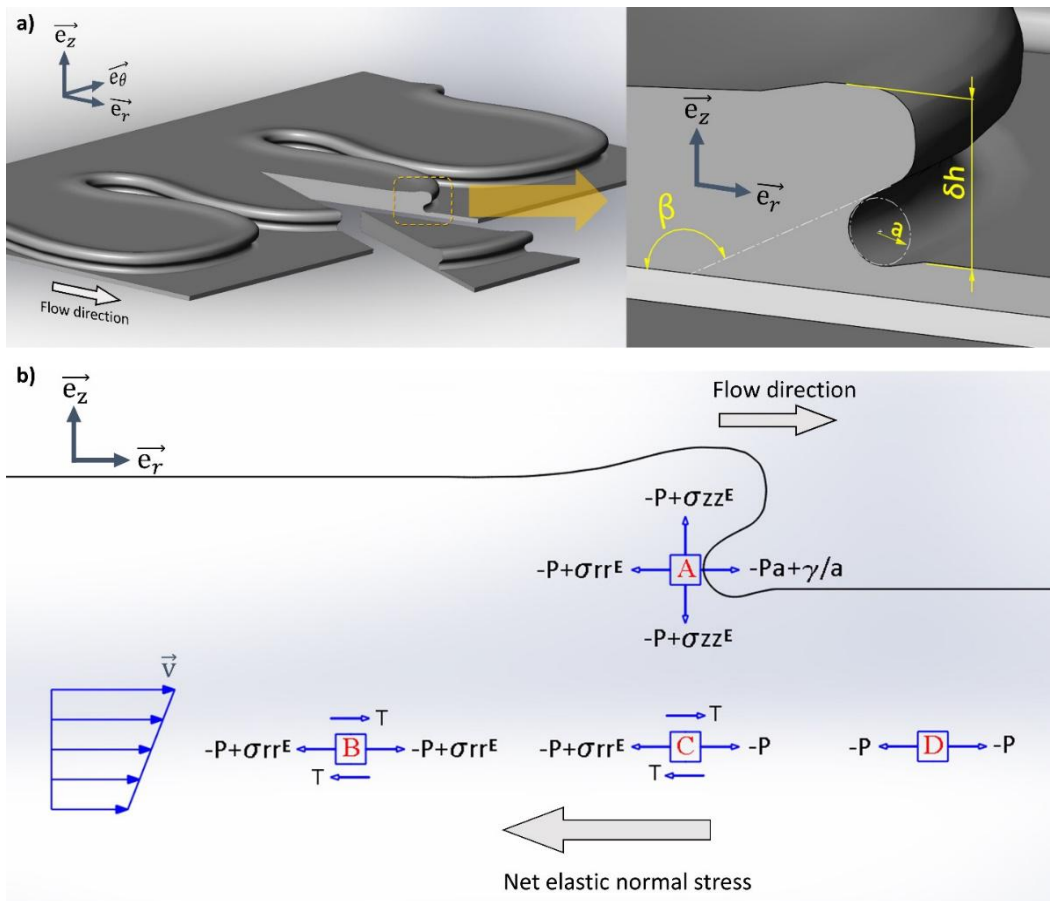

**Figure 4: Sketch of the free surface geometry near the leading edge. a,** The first normal stress difference at high shear rate induces a curvature with radii a at the leading edge, thus folding of the free surface, and an high apparent contact angle β, that spreading of a Newtonian fluid does not exhibit. **b,** state of stress in the upstream of the leading edge (point B), at the leading edge (point C), and downstream (point D) of the leading edge beneath the free surface, and within the viscous shear layer. At large shear rate, the imbalance of normal stress in radial direction at point C results in $\sigma_{rr}^E$ as net normal stress pointing to $-\vec{e_r}$.

**Fingering Instability.** A number of fingers already form in the first several ms of spreading, and the finger number evolves with time (Fig. 5b and 5c). One finger eventually separates into two fingers; It also arrives that one finger loses the competition with its neighbors, and disappear later (Supplementary Fig. 11). We define the average wavelength of the fingers as $\lambda(t) = \frac{2\pi R(t)}{N(t)}$, with $R$ the radius of the leading edge, and $N$ the number of fingers at the moment. In Fig. 5 we show temporal variation of $N$ and λ, for samples with three different Mw and each for different concentrations.

Fig. 5b shows finger number for PEO Mw=5 MDa solution at different concentrations. For each concentration, 3-4 experiments are done to study the distribution of the finger numbers, we find that within 100 ms, the finger number varies strongly from one experiment to another for a fixed concentration, although the spreading radius is reproducible. In addition, within 100 ms, there is no significant difference of finger number between different concentrations. However, later than 100 ms, the finger number in different experiments for a fixed concentration converge to a narrower range of value, and begins to differentiate for different concentrations, solution with higher concentration finally selects fewer number of fingers.

Finger number also depends on molecular weight. Fig. 5c shows finger number for three different Mw: 8 MDa, 5 MDa and 2 MDa. For each Mw, the concentration is chosen for the three samples having similar viscosity (Supplementary Fig. 7). At a given time such as 158.12 ms, the three samples have comparable spreading radius (Fig. 5a), although the one of 8 MDa spreads larger, due to its slightly smaller viscosity compared with the other two samples. Finger number increases with decreasing molecular weight (Fig. 5c), which can also be observed visually in Fig. 5a.

Temporal evolution of the average wavelength $\lambda$ for three different Mw and each for different concentrations are plotted in Fig. 5d and 5e (zoomed version of Fig. 5d, in a shorter time interval). We observe that the wavelength for a fixed sample increases with time, principally due to the fact that once fingers are formed at early time, the wavelength increases when the leading edge enlarges during spreading. Particularly, from Fig. 5d and 5e, if we fix a time and compare among different samples, $\lambda$ increases with increasing Mw, and decreases with increasing concentration.

**Discussions**

During Marangoni spreading on the deep layer, a shear flow is generated within a thin layer beneath the free surface. Due to transient nature, the upstream of the leading edge is under shear effect, whereas the downstream is immobile until the leading edge passes through it. This procedure has been illustrated by PTV[29]. In our case where the fluid possesses viscoelastic property, upstream of the leading edge subjected to strong shear, generates elastic normal stress $\sigma_{rr}^E$, whereas the downstream does not. Thus an imbalance of normal stress occurs in the shear layer, beneath the leading edge, at the point C in Fig. 4b. The resultant normal stress directs to $-\vec{e_r}$, we propose that it is responsible for the formation of fingers. This net elastic normal stress is consistent with the restoring force generated by elastic molecules under abrupt elongation[37].

The restoring force causes swelling in the directions orthogonal to that of filament stretching[38], so that in our case swelling in directions $\vec{e_\theta}$ and $\vec{e_z}$, which can be visually perceived in Supplementary Movie 2. The morphology of our fingering instability is reminiscent of the instability on the free surface of soft elastic materials under swelling, the crease, which was studied theoretically and experimentally[40,41]. However In our work, considering the transient procedure of spreading, we do not minimize the elastic energy to find the wavelength with thermodynamically stable state[41]. We take the first normal stress difference $N_1$ as the characteristic stress which generates the fingers, with elastic effect (elastic modulus $G$) opposing the deformation of the leading edge, we look at their effect on the wavelength selection.

We consider a characteristic strain in the vicinity of the leading edge $\epsilon \sim \frac{N_1}{G} = \frac{N_1 \tau}{\eta}$. In our case with large deformation, where the displacement is larger than the reference length, the relation between the strain and the displacement can be deduced from Lagrangian finite strain tensor[42], that $\varepsilon_{ij} \approx \frac{1}{2}\sum_{k=1,2,3} \frac{\partial u_k}{\partial x_i} \cdot \frac{\partial u_k}{\partial x_j}$, $u$ denotes the displacement, $i, j$ denotes indices of coordinates. Since the displacement in stretching direction is much higher than that in other two directions, we have

$\varepsilon_{rr} \sim \frac{1}{2}\left(\frac{\partial u_r}{\partial r}\right)^2$. The displacement $u_r$ can be described by a characteristic length L, telling the extent of deformation. Thus we have strain $\varepsilon_{rr}$ proportional to $L^2$, so that L is proportional to $(\frac{N_1 \tau}{\eta})^{1/2}$. The extent of leading edge deformation in $\vec{e_r}$ is comparable with the wavelength λ in $\vec{e_\theta}$ (Fig. 1b), so that we have λ proportional to $(\frac{N_1 \tau}{\eta})^{1/2}$.

$N_1$ and η are time dependent, we plot $\left(\frac{N_1 \tau}{\eta}\right)^{\frac{1}{2}}$(t) in Fig. 5f. It is reasonable to observe that this quantity decreases with time, because $N_1$ decreases and η increases during time. We put apart the temporal evolution, and compare at a fixed time among different Mw and concentrations. $\left(\frac{N_1 \tau}{\eta}\right)^{\frac{1}{2}}$ increases with increasing Mw, and decreases with increasing concentration, this is coherent with wavelength measurements shown in Fig. 5d and 5e. We take the moment at 20 ms (close to the early period of spreading), and plot the wavelength λ vs. $\left(\frac{N_1 \tau}{\eta}\right)^{\frac{1}{2}}$ in Fig. 5g. $\left(\frac{N_1 \tau}{\eta}\right)^{\frac{1}{2}}$ clearly reflects the trend of λ variation for different Mw and concentrations. In addition, λ has a power dependence on $\left(\frac{N_1 \tau}{\eta}\right)^{\frac{1}{2}}$ with power=0.1758. This power value varies from one time to another. The experimental result does not show a linear dependence of λ on $\left(\frac{N_1 \tau}{\eta}\right)^{\frac{1}{2}}$, principally because of the transient nature of the spreading, with stress state varies constantly during time. Meanwhile, strain state at a certain moment is tightly affected by stress state in the history[39]. However, here we arrive to describe the trend of wavelength variation for different Mw and concentration, by reasoning on the interplay of elastic normal stress and the elastic modulus.

Despite our scaling analysis does not predict the power value of the wavelength λ dependency on $\left(\frac{N_1 \tau}{\eta}\right)^{\frac{1}{2}}$ in our transient spreading case, we apply this explanation to a similar situation in static state, that unidirectional coating of a viscoelastic polymer film on a solid surface by a blade[15]. In the work of Deblais *et al.*, they observed that the characteristic length of the patterns is proportional to the velocity. They provided an explanation based on characteristic elongational rate of the polymer filaments, which fixes the ratio between characteristic length and velocity as a constant. Here, we explain their observation in terms of elastic normal stress, that characteristic length L is proportional to $\left(\frac{N_1 \tau}{\eta}\right)^{\frac{1}{2}}$. Viscoelastic time τ is a fixed value for a polymer solution, thus in their situation the L is proportional to $\left(\frac{N_1}{\eta}\right)^{\frac{1}{2}}$. Since for a flow under continuous shear[38], $N_1 = 2\eta\tau\dot{\gamma}^2 = 2\eta\tau(\frac{V}{h})^2$, with $V$ the characteristic velocity, and $h$ the film thickness which is measured to be constant in their experiment, thus $N_1$ is proportional to $\eta \cdot V^2$, so that the $\left(\frac{N_1}{\eta}\right)^{\frac{1}{2}}$ is proportional to V, thus the characteristic length L is proportional to the velocity V, which agrees with their observation. Therefore, we think that our reasoning based on $\left(\frac{N_1 \tau}{\eta}\right)^{\frac{1}{2}}$ captures the dominant effect for wavelength selection, which is the first normal stress difference in strong shear flow.

Lastly we want to discuss the effect of shear thinning of polymer solution. We made Boger fluids[43] with viscoelastic property, however show negligible shear thinning effect. The rheological property of the Boger fluids are shown in Supplementary Fig. 5, viscosity and $N_1$ values are comparable with the shear thinning PEO solutions. These Boger fluids do not show finger formation during Marangoni spreading, but only show a distinguishable leading edge (Supplementary Movie 3). This effect indicates that shear thinning is another important factor for finger formation. However we think that the nonlinear elastic effect is still important for wavelength selection, since Boger fluids do form fingers at the contact line of thin film flow on an inclined solid surface (Supplementary Fig. 12). Shear thinning has been shown to play an important role in amplification of velocity perturbation[38,44]. Here we propose a mechanism that shear thinning may have in the amplification of the leading edge perturbation. Suppose at a fixed time $t$ that a wavy perturbation is preset along the circumferential direction of the leading edge. In this shear flow of a shear thinning fluid, an infinitesimal higher velocity leads to higher shear rate, since we consider the thickness of the viscous shear layer is constant along circumferential direction. Thus positions which are more severely sheared have lower viscosity. Under constant Marangoni stress, lower viscosity causes higher velocity, therefore a perturbation of velocity along the leading edge is amplified by this positive feedback mechanism, and forms fingers. The logic of this mechanism is reminiscent of that of viscous fingering in Saffman-Taylor instability[14,45,46]. However this fingering instability in Marangoni spreading is qualitatively different from Saffman-Taylor instability, since in our spreading case the wavelength selection is dominated by the elastic normal stress, instead of viscosity contrast.

As a summary, we rely on the nonlinear elasticity under strong shear flow, to explain the wavelength selection during Marangoni spreading on the free surface of a deep layer of polymer solutions. Due to the transient nature of Marangoni spreading, and the stress history dependency of the strain state of viscoelastic fluids, dynamic procedure of wavelength selection is out of the scope of this study. We hope this experimental work can stimulate theoretical and numerical work considering transient growth of fingers[26,27] and constitutive equation of complex fluids. This work may have important implications in treating airway infection diseases using low cost and easily manipulated inhalation methods, through atomization of drug components.

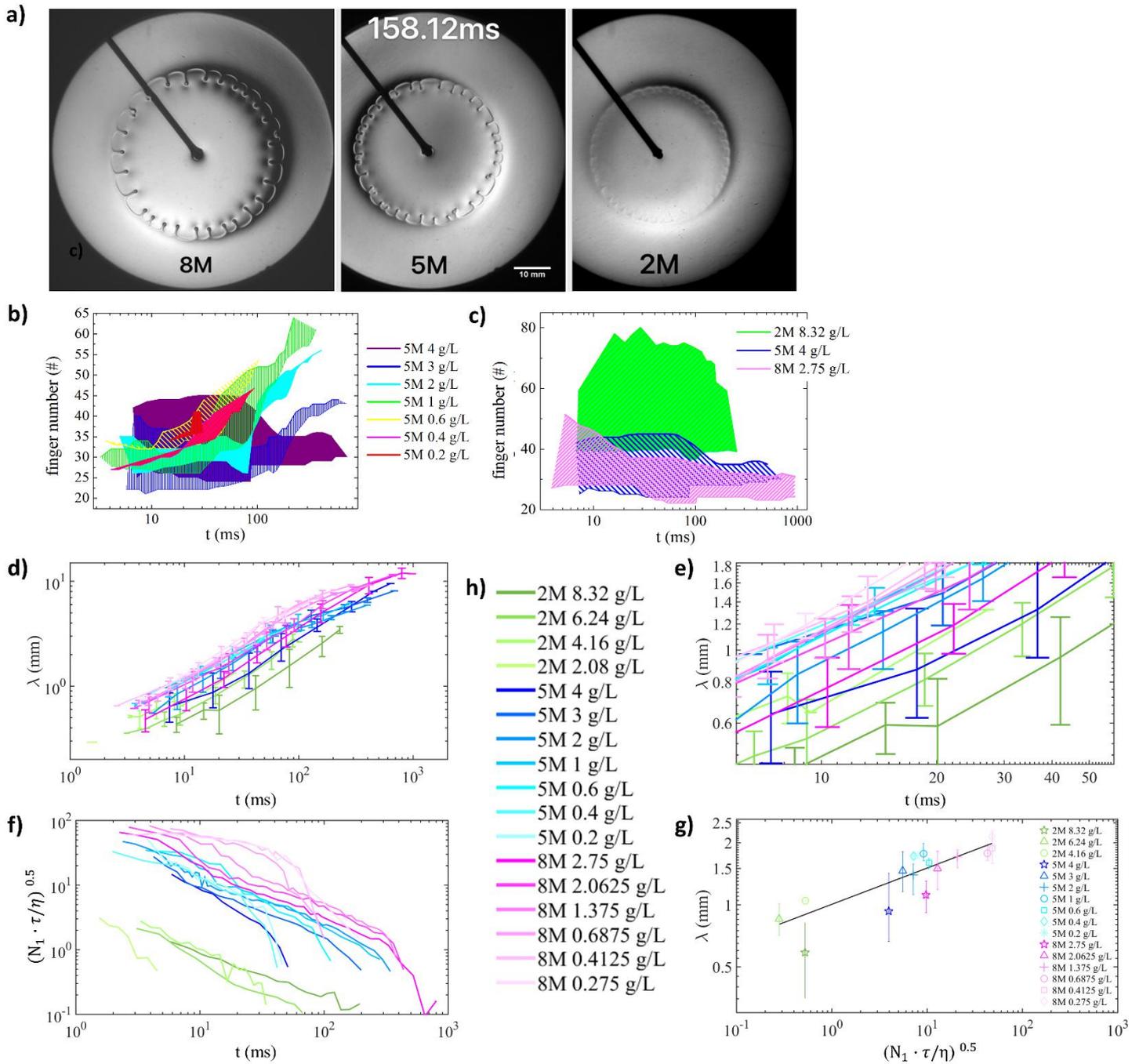

**Figure 5: Fingering instability and wavelength. a,** Spreading pattern at 158.12 ms, for PEO solution of 8 MDa at 2.75 g/L, 5 MDa at 4 g/L, and 2 MDa at 8.32 g/L. **b,** Finger number vs. time, for Mw=5 MDa PEO solution at different concentration. **c,** Finger number vs. time for solutions with three different Mw shown in **a**. **d,** Temporal evolution of the wavelength for three different Mw and each with different concentration. **e,** Zoomed version of figure **d** within smaller time interval for clear comparison of the wavelength among different samples. **f,** Temporal evolution of $\left(\frac{N_1 \tau}{\eta}\right)^{\frac{1}{2}}$ for all samples. **g,** Correspondence between wavelength $\lambda$ and $\left(\frac{N_1 \tau}{\eta}\right)^{\frac{1}{2}}$ at t = 20 ms. **h,** Legend for figure **d**, **e** and **f**.

## Methods

**Formulation of solutions.** Polyethylene oxide (PEO) powder (Sigma-Aldrich) with different molecular weight (300 KDa, 2 MDa, 5 MDa, 8 MDa) and Polyethylene glycol (PEG) powder (Sigma-Aldrich) are dissolved in deionized water (resistivity=18.2 MΩ, Shanghai Titan Scientific Co.), under stirring at 30 rpm at room temperature. Solutions are made for different concentrations for each molecular weight. Boger fluids are made by adding small amount of large molecular weight PEO into high viscosity PEG solution. Powder is added into a prewetted glass bottle, forms an evenly distributed thin layer of powder, then add DI water. This method avoids agglomeration of PEO powder in water, so that reduces time for dissolution. Bottle cap is sealed to prevent polymer chain denaturation. Completely dissolved solution is stored in refrigerator, to be used within days, to avoid experiments being influenced by thinning and denaturation during time[47]. Non-ionic surfactant TritonX-100 (Beijing Solarbio Science & Technology Co.) is dissolved in DI water, and added to the drop phase. For instance, 20 cmc surfactant aqueous solution is mixed with the 4 g/L PEO solution with same volume, to form a sample of 2 g/L PEO solution with 10 cmc surfactant.

**Imaging technique.** For vertical view experiments, both reflection and transmission optical methods are used for observation (Supplementary Fig. 2). A Xenon light (HDL-II 470VA, Suzhou Beitejia Photoelectric Technology Co) is used to illuminate the phenomenon. In reflection method, a white paper is placed under the bath container, light source and camera are both on the top of the free surface. Reflection method allows the observation of large view up to 15 cm diameter, however with less resolution for the small spreading radius at very beginning of the phenomenon. Therefore, the transmission method is applied, with light illuminating the bath from top, beam diameter been enlarged by a lens, and be registered by the camera on the bottom. Transmission method supplements the reflection method by allowing high spatial resolution for the phenomenon at the beginning. Results of the two methods are proved to be coherent with each other. For side view experiments, a LED panel (OPT Machine Vision Tech Co., model OPT-DPA1024E-4, output power 48 W) illuminates the liquid surface, with camera focused to the dropping position. The spreading dynamics is captured by fast camera (PCO. dimax HS1) at 7039.08 fps, with 1000 * 1000 pixels, and dynamical range 16 bits.

A drop of polymer solution with surfactant is deposited on a deep layer of polymer solution without surfactant, the bath solution is 8 mm deep, in a square container made of Plexiglas with side length 22.5 cm for reflection method, and a square petri dish with side length 12 cm for transmission method. The drop falls from 4 mm above the free surface. Drop volume varies from 12.4 μl to 12.8 μl for 2 MDa PEO solutions, and 13.9 μl to 15.7 μl for 8 MDa PEO solutions, freed by a syringe pump (Harvard 11 Pico Plus Elite); and from 10.5 μl to 17.75 μl for 5 MDa PEO solutions, freed manually. It is proved that a doubling of drop volume does not introduce significant variation on wavelength. Experiments are performed at temperature $297 - 301\ K$.

**Rheometry.** Rheological properties of the solutions are measured by a Rheometer Anton-Paar MCR 302, rotor CP50-1, with 50 mm diameter and cone angle of $1°$. Viscosity and first normal stress are measured in the shear rate range from 0.1 to 10000 $s^{-1}$, at temperature 298 K.

**Transmission Lattice Method.** Deformation of the free liquid surface is measured by Transmission Lattice Method[48]. A transparent sheet with black grids is placed under the transparent container, and the LED light for illumination is under the sheet. Grids frequency on the sheet is 5 lines per mm. Observing the grids above the deformed liquid surface, the grid pattern is distorted by the modulation of the liquid surface topography. Horizontal displacement of the grids pattern caused by liquid surface perturbation is analyzed by Sampling Moiré method combined with the phase-shift technique[49]. Fast camera records the temporal deformation of grids pattern. Taking the unperturbed and known liquid height far away from the drop deposition area as starting point of calculation, the surface height perturbed by the drop can be iterated from the infinite to the closed area using Newton iterative algorithms. So that during image recording, it is necessary to include the unperturbed area in each frame. Meanwhile, high resolution is also necessary for precision on the determination of deformation. Therefore, compromises between view scope and resolution should be made. Using this technique, surface deformation is measured for the early period (about the first 10-20 ms) with theoretical sensitivity better than 1 μm, when height perturbation is well detectable.

**Data availability.** All relevant data presented in the paper is available upon request to the corresponding author.

## Acknowledgements

We acknowledge H.A. Stone, QQ. Liu, M-J. Thoraval, GK. Hu, CJ. Lv, K. Zhang, M. Roché, TT. Kong, YN. Liu, K. Luo, C. Wan, XL. Wang and XF Huang for helpful discussions. We acknowledge HS. Chen, J. Li, XD. Chen, H. Liang, and Y. Yu for help on experimental equipment. ZZ. Li acknowledge the BIT Young researcher starting program, China NSFC, grant no. 11802023, the Shanghai Sailing Program, grand no. 17YF1409200, ZW. Liu acknowledge China NSFC, grand no. 11572041 for financial support.

## Author contributions

X.M. carried out the experiments with the help of ZZ.L.; X.M., ZW.L. and ZZ.L. designed the experiment. All authors contributed to data analysis and interpretations. X.M. and ZZ.L. wrote the paper with contributions from all authors.

## Additional information

Supplementary information accompanies this paper is provided.

**Competing financial interests:** The authors declare no competing financial interests.

# Supplementary information

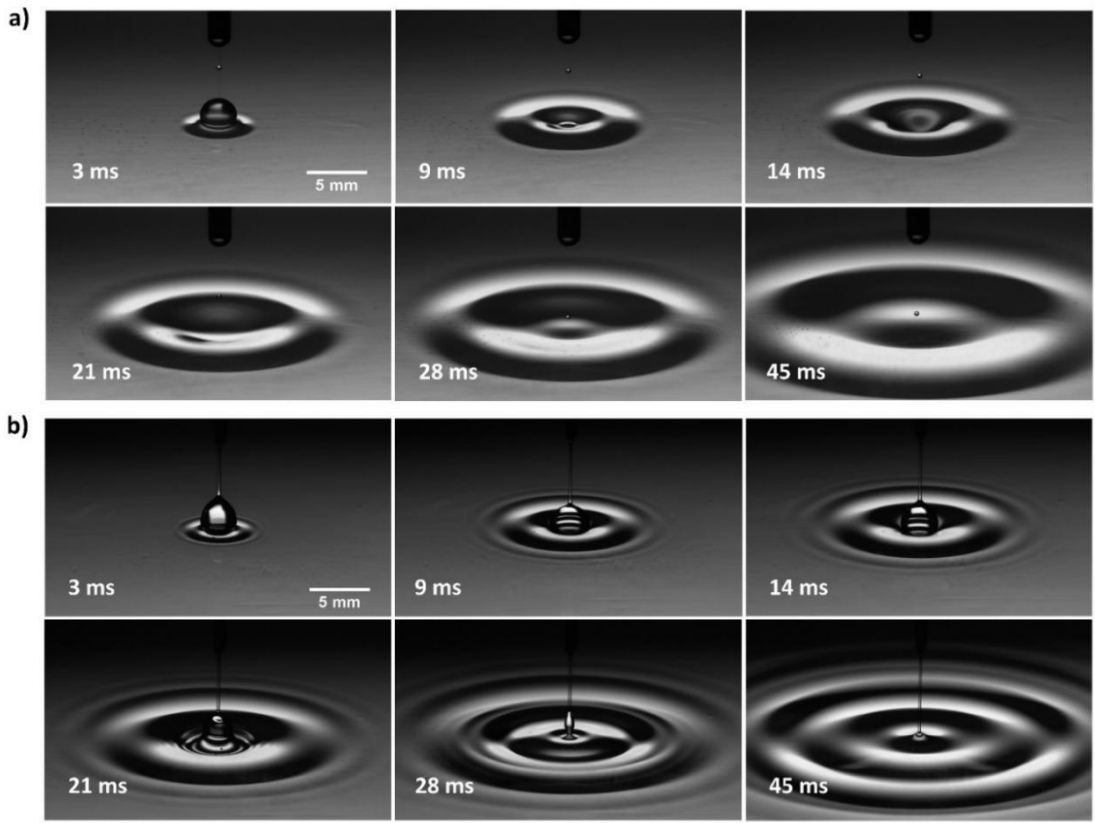

**Supplementary Figure 1: Merging and spreading of a drop on the free liquid surface. a,** A drop of a high molecular weight PEO solution (Mw=5 MDa, C=4 g/L, shear thinning and viscoelastic) without surfactant, merging with the deep layer (8 mm) of the same fluid. Delayed merging and the propagation of surface wave during the merging are observed, no fingering instability is observed. **b,** A drop of a low molecular weight PEO solution (Mw=300 KDa, C=20 g/L, Newtonian fluid $\eta = 37.5$ mPa·s) with 10 cmc of Triton X-100 non-ionic surfactant spreading on the deep layer (8 mm) of the same PEO solution without surfactant. Merging and spreading occur instantaneously, spreading does not lead to the fingering instability, and the leading edge is not distinguishable.

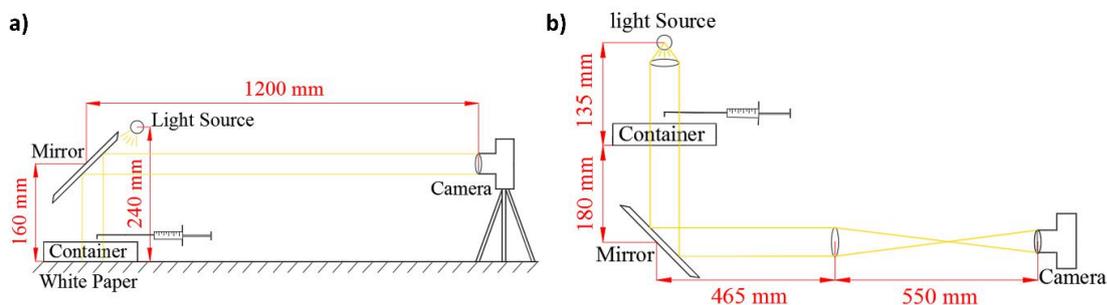

**Supplementary Figure 2: Light reflection and transmission method. a),** Reflection method by putting a white paper under the container and register from above, allow larger view scope but less spatial resolution at the beginning of spreading. **b),** Transmission method filming from the bottom, allow better spatial resolution at the beginning of spreading.

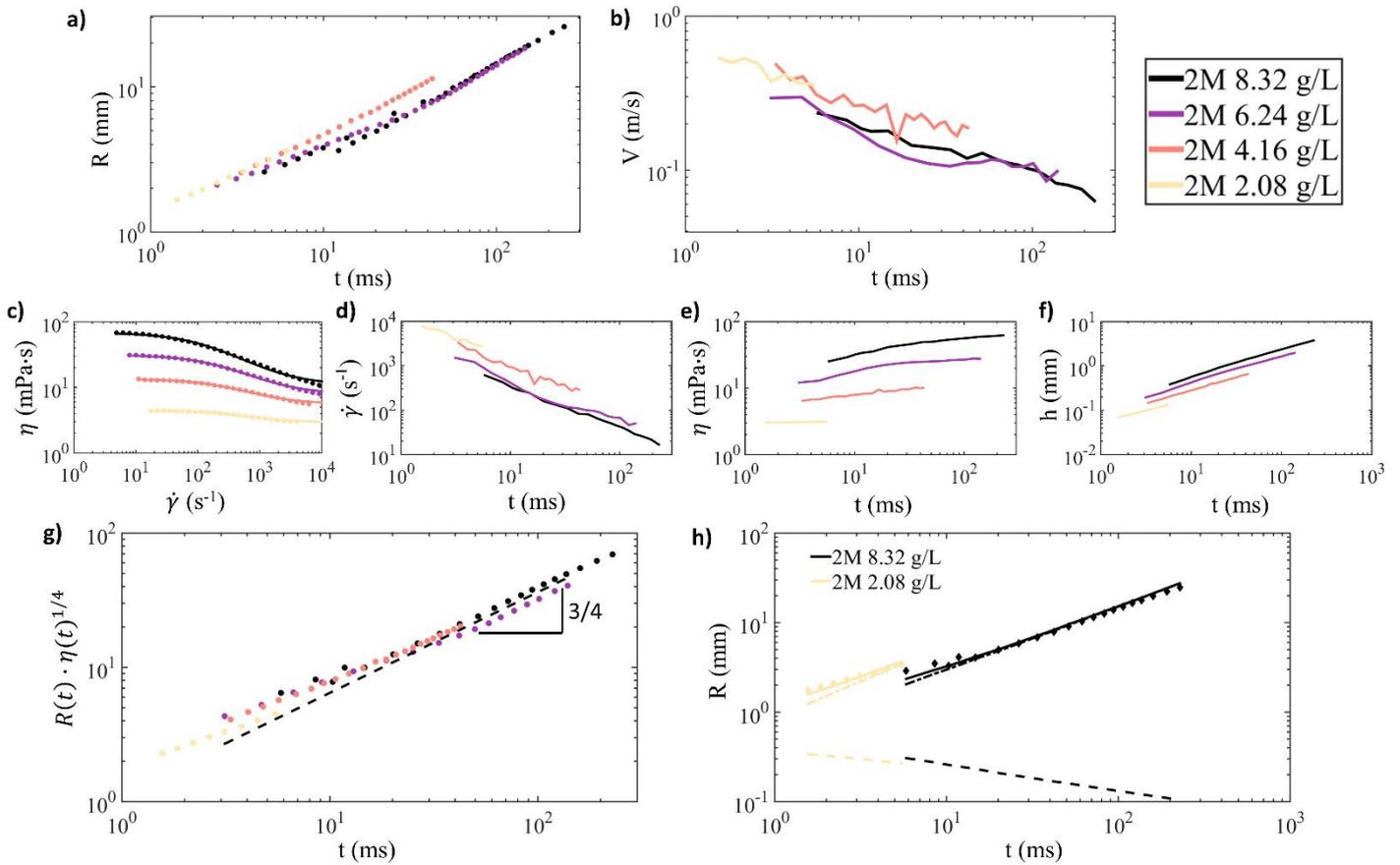

**Supplementary Figure 3: Spreading radius of PEO drop with surfactant on the deep layer of PEO solution without surfactant, Mw=2 MDa for different concentration. a,** Experimentally measured spreading radius as function of time. Each solution is tested using method of transmission and reflection, both coherent with each other. **b,** Velocity of the leading edge vs. time, deduced from differentiating the radius. **c,** Viscosity vs. shear rate measured by a rheometer, and fitted by the Cross model. **d,** Shear rate vs. time calculated from equation (1). **e,** Viscosity vs. time for shear thinning fluids. **f,** Thickness of the viscous diffusion layer vs. time. **g,** Temporal evolution of $R(t) \cdot \eta(t)^{1/4}$ for different concentration. **h,** Radius measured in experiment (diamond) agrees with the calculated radius $r_e + r_\eta$ from equation (2) (solid lines), with a multiplicative factor 1.35. Two samples of PEO 2 MDa at 8.32 g/L (black) and 2.08 g/L (yellow) are shown, dashed lines for the elastic displacement $r_e(t)$, and dash-dotted lines for the viscous displacement $r_\eta(t)$. $r_\eta$ dominates over $r_e$ for the whole procedure.

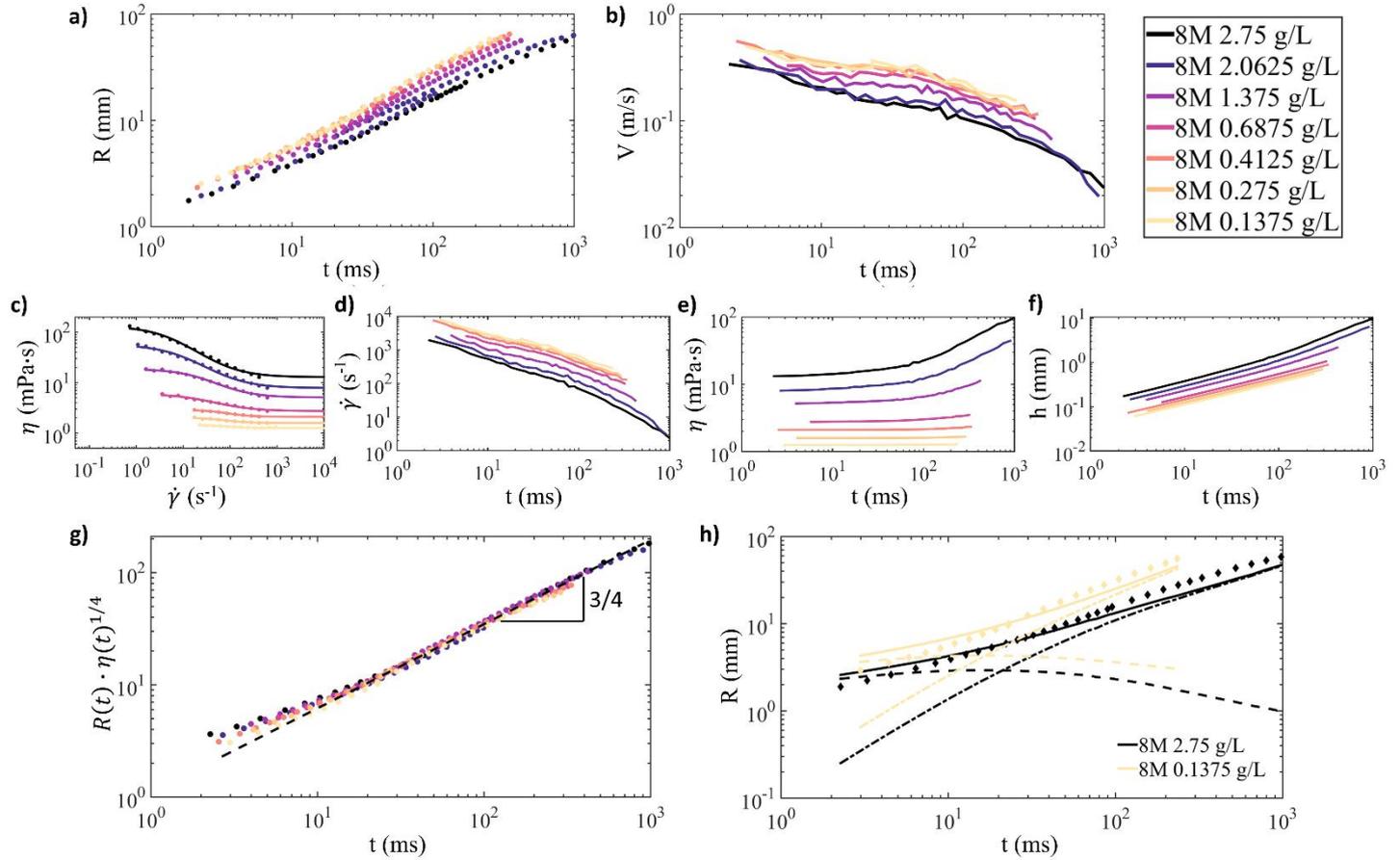

**Supplementary Figure 4: Spreading radius of PEO drop with surfactant on the deep layer of PEO solution without surfactant, Mw=8 MDa for different concentration. a,** Experimentally measured spreading radius as function of time. Each solution is tested using the method of transmission and reflection, both coherent with each other. **b,** Velocity of the leading edge vs. time, deduced from differentiating the radius. **c,** Viscosity vs. shear rate measured by a rheometer, and fitted by the Cross model. **d,** Shear rate vs. time calculated from equation (1). **e,** Viscosity vs. time for shear thinning fluids. **f,** Thickness of the viscous diffusion layer vs. time. **g,** Temporal evolution of $R(t) \cdot \eta(t)^{1/4}$ for different concentration. **h,** The radius measured in experiment (diamond) agrees with the calculated radius $r_e + r_\eta$ from equation (2) (solid lines) with a multiplicative factor 0.85. Two samples of PEO 8 MDa at 2.75 g/L (black) and 0.1375 g/L (yellow) are shown, dashed lines for the elastic displacement $r_e(t)$, and dash-dotted lines for the viscous displacement $r_\eta(t)$. $r_\eta$ and $r_e$ interchange importance at $t = \tau$.

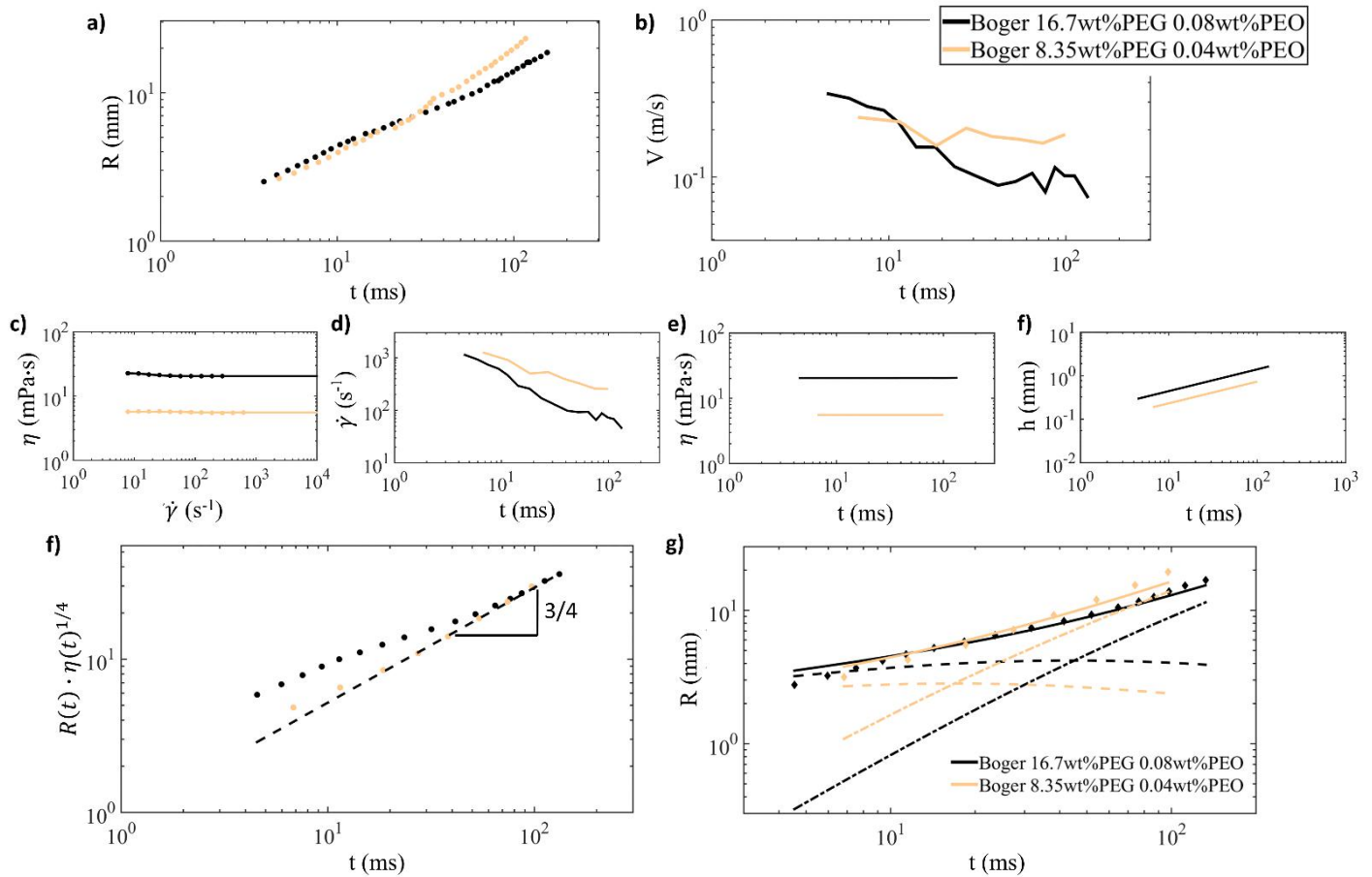

**Supplementary Figure 5: Spreading radius of drop with surfactant on the deep layer of solution without surfactant for two Boger fluids (16.7wt% PEG 8KDa+0.08wt% PEO 5MDa, and 8.35wt% PEG 8KDa +0.04wt% PEO 5MDa). a,** Experimentally measured spreading radius as function of time. Each solution is tested using method of transmission and reflection, both coherent with each other. **b,** Velocity of the leading edge vs. time, deduced from differentiating the radius. **c,** Viscosity vs. shear rate measured by a rheometer, and fitted by the Cross model. **d,** Shear rate vs. time calculated from equation (1). **e,** Viscosity vs. time for shear thinning fluids. **f,** Thickness of the viscous diffusion layer vs. time. **g,** Temporal evolution of $R(t) \cdot \eta(t)^{1/4}$ for different concentration. **h,** The radius measured in experiment (diamond) agrees with the calculated radius $r_e + r_\eta$ from equation (2) (solid lines) with a multiplicative factor 0.75. Dashed lines for the elastic displacement $r_e(t)$, and dash-dotted lines for the viscous displacement $r_\eta(t)$. $r_\eta$ and $r_e$ interchange importance at t = τ.

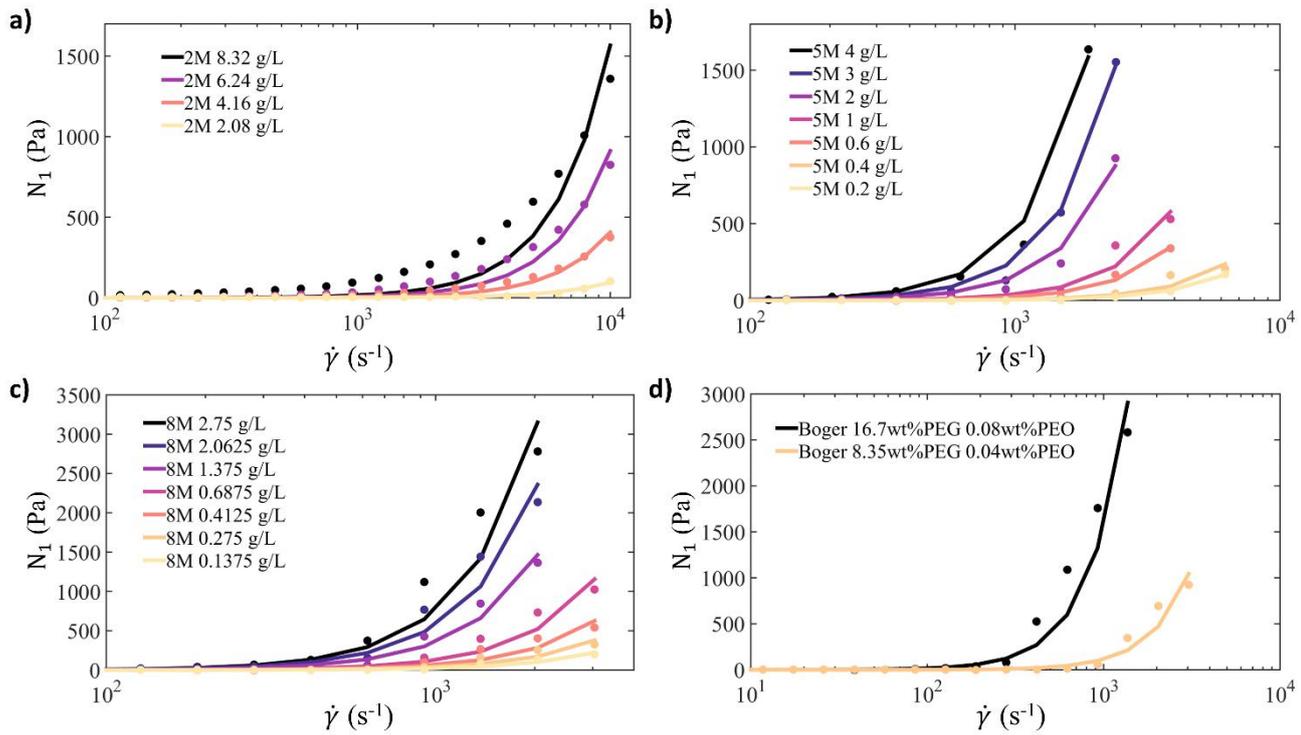

**Supplementary Figure 6: First normal stress difference $N_1(\dot{\gamma})$ measured by a rheometer.** We obtain the viscoelastic relaxation time $\tau$ by fitting the experimentally measured $N_1(\dot{\gamma})$, based on the relation that $N_1 = \Psi_1 \dot{\gamma}^2$, with $\Psi_1 = 2nk_BT\tau^2$ the coefficient of first normal stress difference, $n$ the number density of polymer molecules, and $k_BT$ the thermal energy. The obtained relaxation times are listed in Supplementary Table 1.

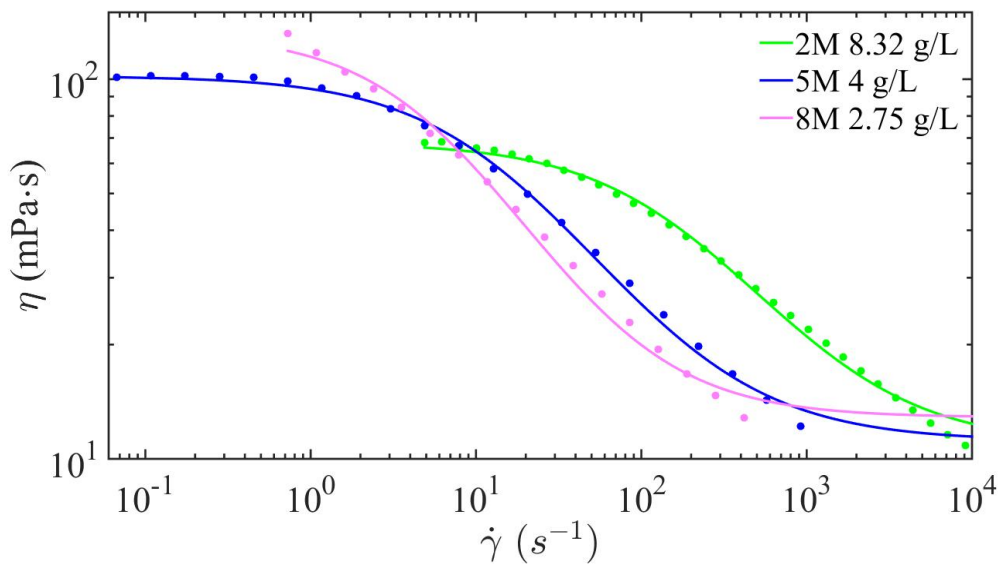

**Supplementary Figure 7: Viscosity of solutions of 8 MDa at 2.75 g/L, 5 MDa at 4 g/L, and 2 MDa at 8.32 g/L.** The three solutions have comparable viscosity, although 2 MDa solution presents higher viscosity, which explains its smaller radius at a fixed time in Fig. 5a.

| Fluid | concentration | Viscoelastic relaxation time (ms) | filament thinning time (ms) |
|---|---|---|---|
| 2M PEO | 2.08 g/L | 0.43073±0.00666 | 31.01±2.49 |
| | 4.16 g/L | 0.62849±0.00776 | 37.16±4.73 |
| | 6.24 g/L | 0.7683±0.01103 | 40.22±2.09 |
| | 8.32 g/L | 0.87206±0.01711 | 76.66±6.19 |
| 5M PEO | 0.2 g/L | 4.67161±0.05504 | 42.51±4.42 |
| | 0.4 g/L | 3.9405±0.1441 | 108.7±12.9 |
| | 0.6 g/L | 6.2486±0.0884 | 121.2±8.85 |
| | 1 g/L | 6.2559±0.1432 | 252.6±17.0 |
| | 2 g/L | 8.8003±0.1165 | 221.2±19.9 |
| | 3 g/L | 9.4557±0.0589 | 354.3±22.2 |
| | 4 g/L | 10.5860±0.1084 | 578.1±48.8 |
| 8M PEO | 0.1375 g/L | 16.7365±0.4256 | 107.5±14.95 |
| | 0.275 g/L | 15.5117±0.4805 | 298.2±29.2 |
| | 0.4125 g/L | 16.2934±0.5144 | 349.7±25.9 |
| | 0.6875 g/L | 17.1453±0.4051 | 451.7±23.8 |
| | 1.375 g/L | 20.3333±0.3194 | 809.3±26.6 |
| | 2.0625 g/L | 21.0689±0.4225 | 852.8±38.6 |
| | 2.75 g/L | 21.0887±0.5009 | 1025.9±47.2 |
| Boger fluid | 8.35 wt% PEG 8 KDa 0.04 wt% PEO 5 MDa | 16.8917±1.078 | Drop:192.1±24.4 layer:118.3±10.5 |
| | 16.7 wt% PEG 8 KDa 0.08 wt% PEO 5 MDa | 44.2097±0.4125 | Drop:881.1±62.6 layer:622.8±57.9 |

**Supplementary Table 1:** Viscoelastic relaxation time fitted from $N_1(\dot{\gamma})$ of rheology (Supplementary Fig. 6), and characteristic filament thinning time (fitted from Supplementary Fig. 8).

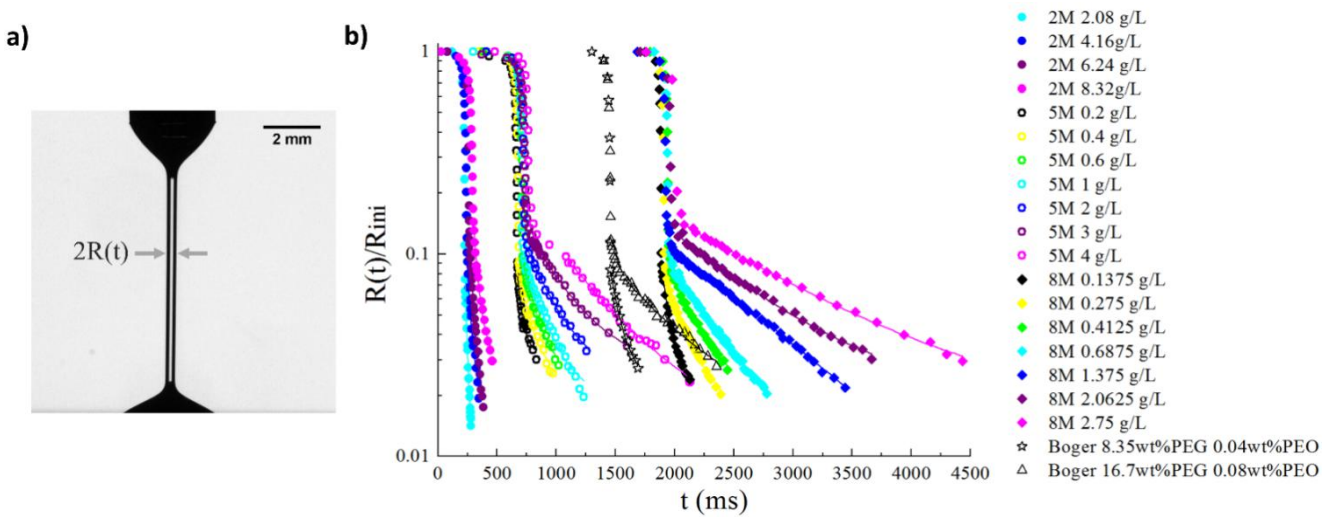

**Supplementary Figure 8: Obtaining the characteristic filament thinning time from the filament thinning experiments. a),** Filament thinning experiment. **b),** The filament radius normalized by the initial radius is measured as function of time. The characteristic filament thinning time of the solutions with Mw=2 MDa, 5 MDa, 8 MDa and Boger fluid are obtained by fitting the curves of pinching dynamics[50,51]. Four sets of curves for different Mw are assembled in one figure, since the obtained characteristic times are invariant by moving the curves along the horizontal axis.

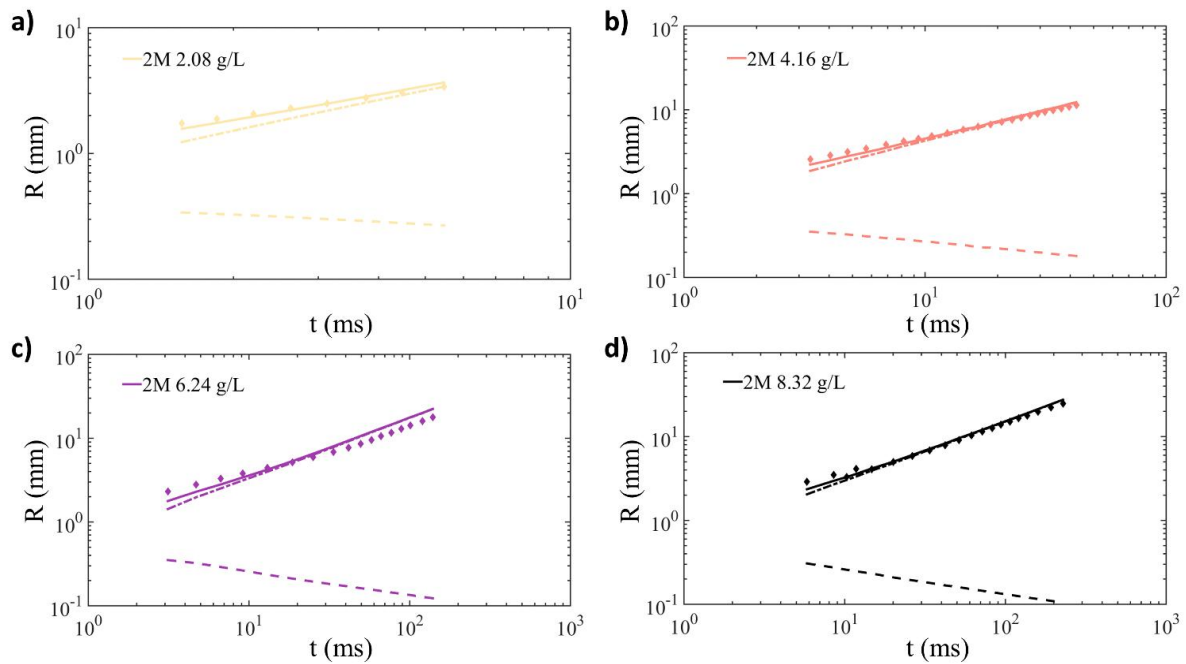

**Supplementary Figure 9-1: Spreading radius of PEO 2 MDa solutions, experiments and theory.** Comparison between experimentally measured spreading radius (diamond) with the radius $r_e + r_\eta$ calculated from Equation (2) (solid line), for different concentration of PEO 2 MDa solution. Dashed line for the elastic displacement $r_e$, and dash-dotted line for the viscous displacement $r_\eta$. The calculated radius (solid line) is multiplied by a factor from 1.3 to 1.35 to be consistent with the experimentally measured radius.

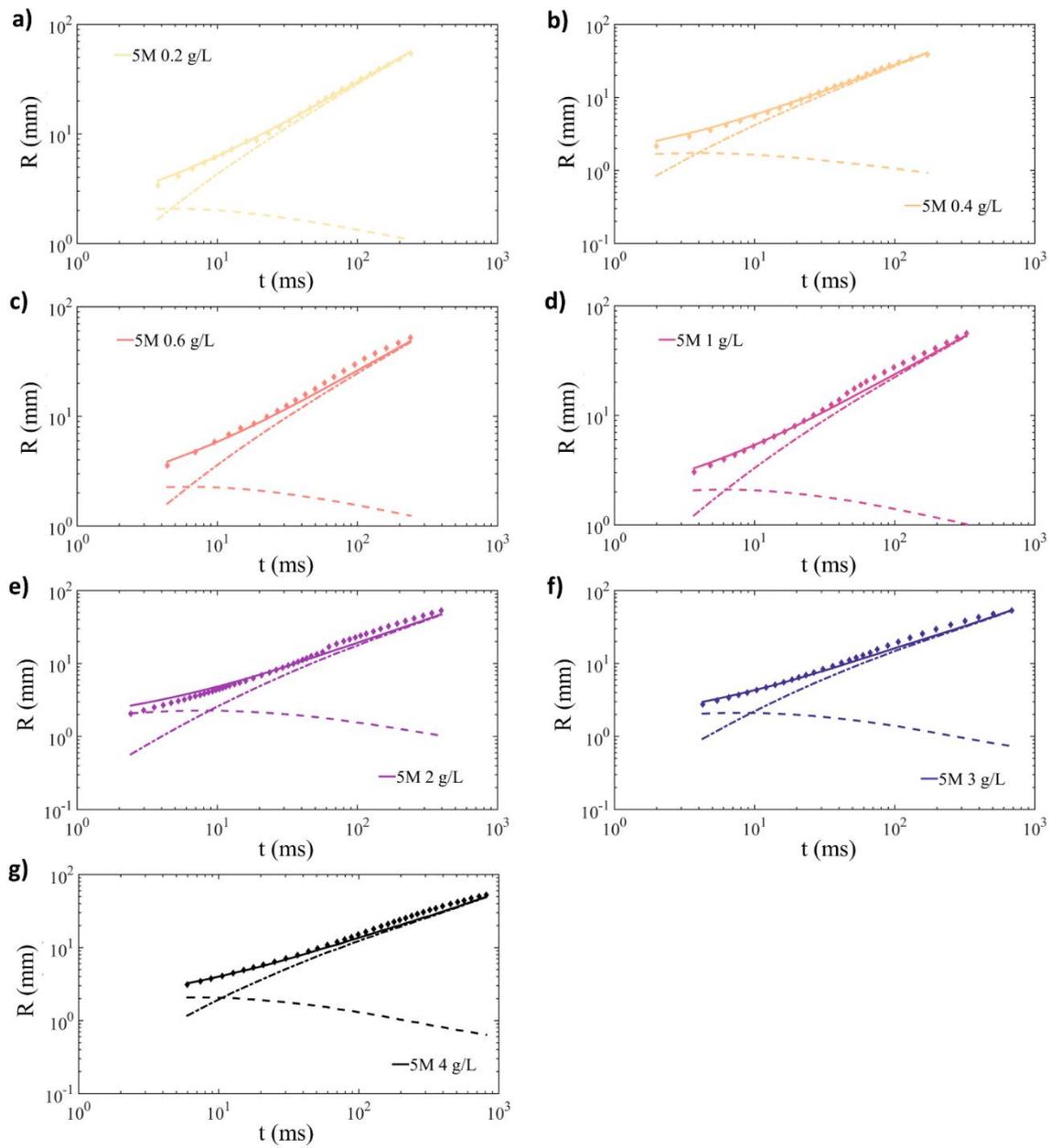

**Supplementary Figure 9-2: Spreading radius of PEO 5 MDa solutions, experiments and theory.** Comparison between experimentally measured spreading radius (diamond) with the radius $r_e + r_\eta$ calculated from Equation (2) (solid line), for different concentration of PEO 5 MDa solution. Dashed line for the elastic displacement $r_e$, and dash-dotted line for the viscous displacement $r_\eta$.

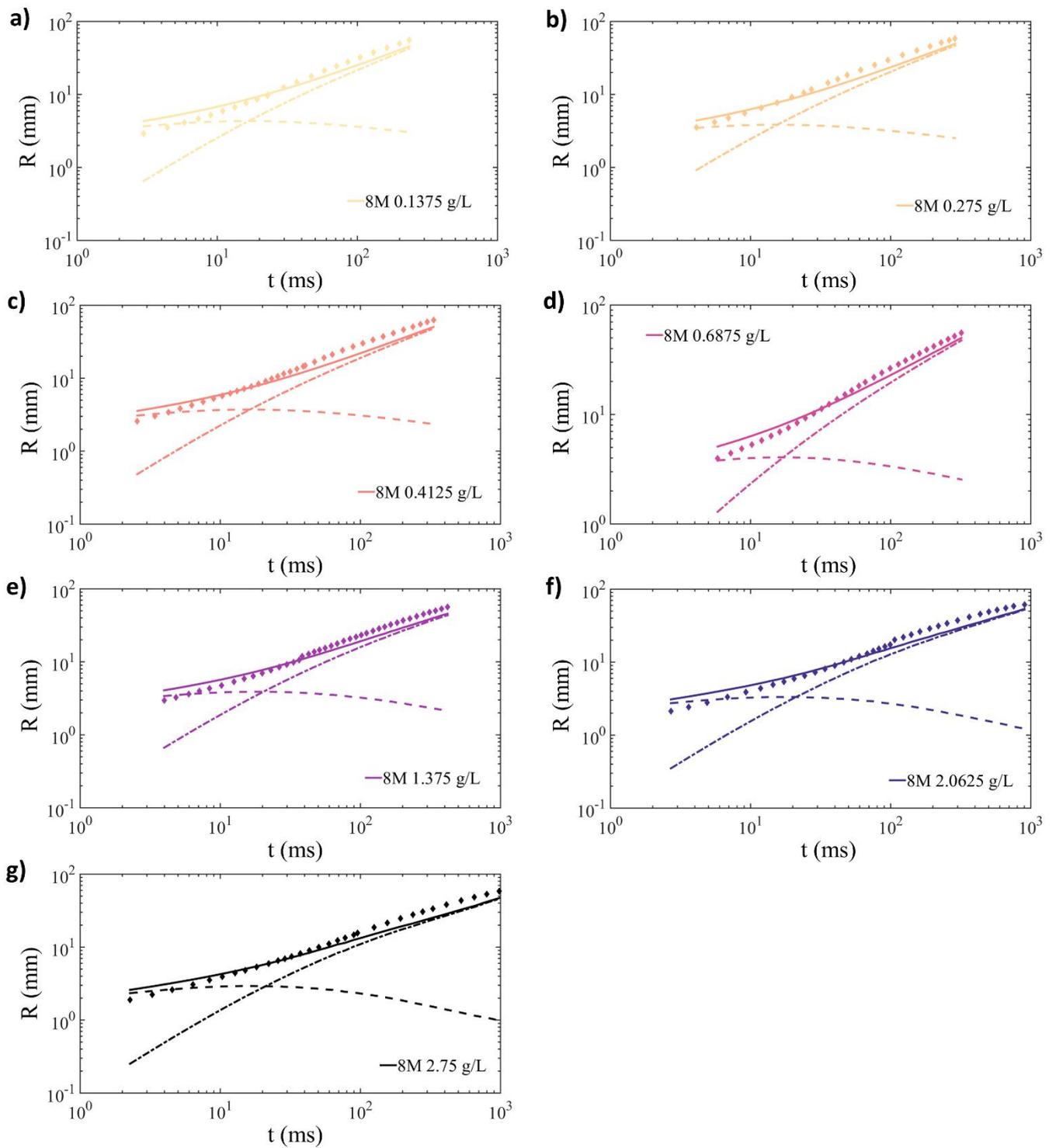

**Supplementary Figure 9-3: Spreading radius of PEO 8 MDa solutions, experiments and theory.** Comparison between experimentally measured spreading radius (diamond) with the radius $r_e + r_\eta$ calculated from Equation (2) (solid line), for different concentration of PEO 8 MDa solution. Dashed line for the elastic displacement $r_e$, and dash-dotted line for the viscous displacement $r_\eta$. The calculated radius (solid line) is multiplied by a factor from 0.8 to 0.9 to be consistent with the experimentally measured radius.

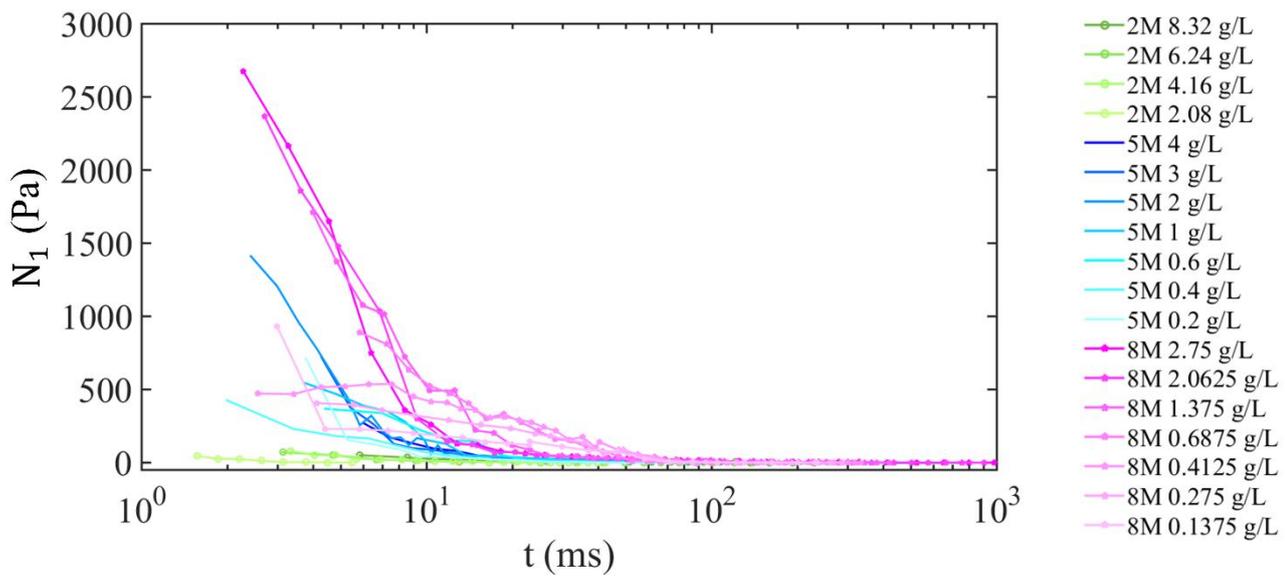

**Supplementary Figure 10: First normal stress difference vs. time**, for three different molecular weights, and each with different concentration.

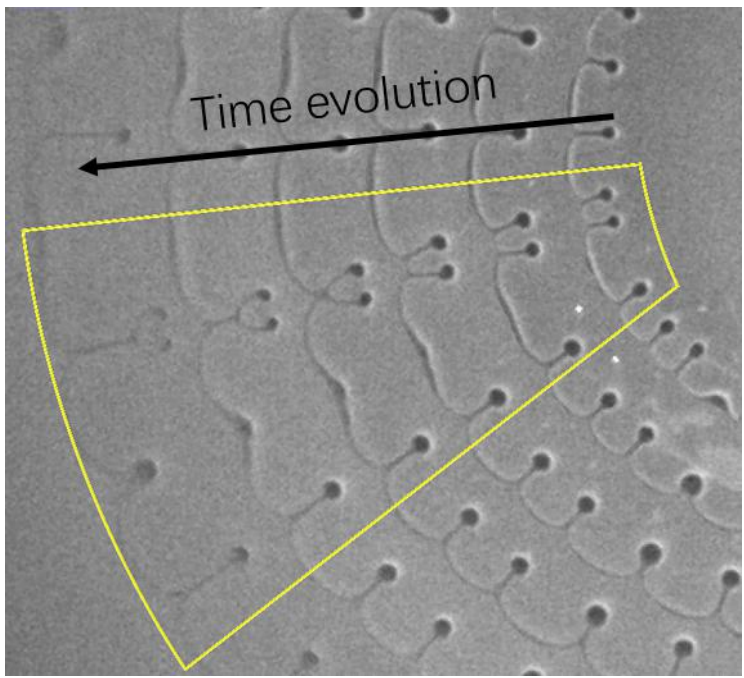

**Supplementary Figure 11: Procedure of new finger formation during spreading, and finger disappearance.** The image shows a superposition of leading edge morphology for a set of time steps.

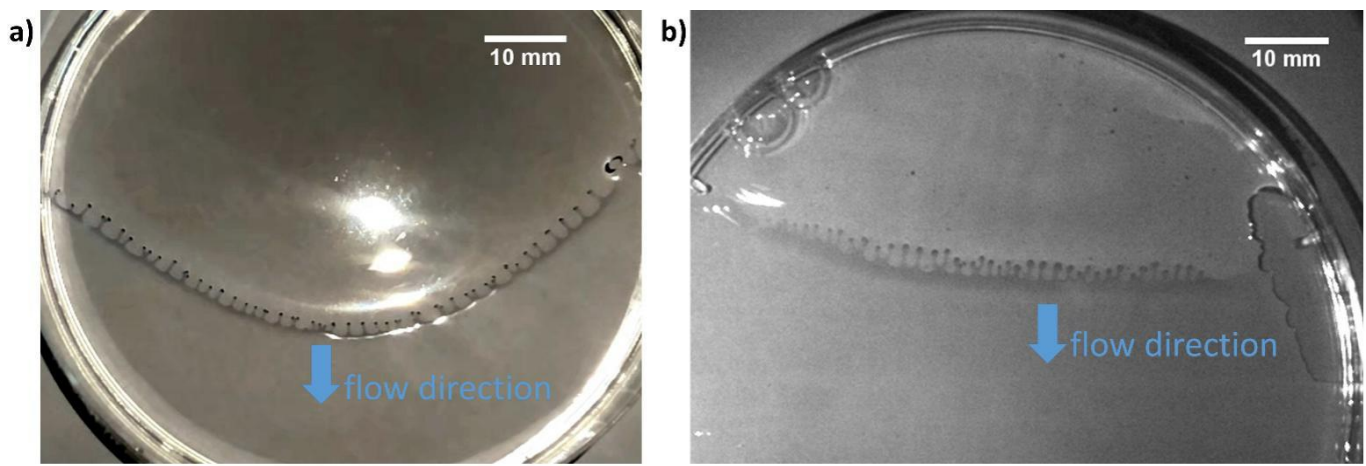

**Supplementary Figure 12: Boger fluid and PEO dilute solution flowing down at an inclined surface prewetted by the same solution. a,** Boger fluid, made of 16.7 wt% PEG 8 KDa + 0.08 wt% PEO 5 MDa. **b,** PEO 8 MDa 1.375 g/L dilute solution. The fingering instability are observed in both fluids.

### Supplementary References

50. Clasen, C., Eggers, J., Fontelos, M. A., Li, J. & Mckinley, G. H. The beads-on-string structure of viscoelastic threads. *J. Fluid Mech.* **556,** 283-308 (2006).
51. Deblais, A., Harich, R., Colin, A. & Kellay, H. Taming contact line instability for pattern formation. *Nat. Commun.* **7,** 12458 (2016).